\documentclass{aa}
\usepackage[varg]{txfonts}
\usepackage{graphicx}
\usepackage{natbib}
\usepackage{subfigure}
\usepackage{afterpage}
\usepackage{amsmath,amssymb,wasysym}
\bibpunct{(}{)}{;}{a}{}{,} 

\newcommand{\grad}{^{\circ}}
\newcommand{\hatap}{\mathrel{\hbox{\rlap{\lower.1ex\hbox{$\approx$}} \kern-.1em \raise1ex \hbox{$\scriptscriptstyle\wedge$}}}}
\newcommand{\lsim}{\mathrel{\hbox{\rlap{\lower.55ex\hbox{$\sim$}} \kern-.3em \raise.4ex \hbox{$<$}}}}
\newcommand{\gsim}{\mathrel{\hbox{\rlap{\lower.55ex\hbox{$\sim$}} \kern-.3em \raise.4ex \hbox{$>$}}}}

\begin{document}

\title{The magnetic field structure of the central region in M31}

\author{R.~Gie\ss\"ubel\inst{1}
\and R.~Beck\inst{1}}

\institute{Max-Planck-Institut f\"ur Radioastronomie, Auf dem H\"ugel 69, 53121 Bonn, Germany\label{1}}

\titlerunning{The Magnetic Field Structure of the Central Region in M31}
\authorrunning{R.~Gie\ss\"ubel \& R.~Beck}

\abstract
{The Andromeda Galaxy (M31) is the nearest grand-design spiral galaxy. Thus far most studies in the radio regime concentrated on the 10~kpc ring. The central region of M31 has significantly different properties than the outer parts: The star formation rate is low, and inclination and position angle are largely different from the outer disk.}
{The existing model of the magnetic field in the radial range $6\leq r\leq14$~kpc is extended to the innermost part $r\leq0.5$~kpc to ultimately achieve a picture of the entire magnetic field in M31.}
{We combined observations taken with the VLA at 3.6~cm and 6.2~cm with data from the Effelsberg 100-m telescope to fill the missing spacings of the synthesis data. The resulting polarization maps were averaged in sectors to analyse the azimuthal behaviour of the polarized intensity ($PI$), rotation measure ($RM$), and apparent pitch angle ($\phi_{\rm obs}$). We developed a simplified 3-D model for the magnetic field in the central region to explain the azimuthal behaviour of the three observables.}
{Our 3-D model of a quadrupolar or dipolar dynamo field can explain the observed patterns in $PI$, $RM$, and $\phi_{\rm obs}$, while a 2-D configuration is not sufficient to explain the azimuthal behaviour. In addition and independent of our model, the $RM$ pattern shows that the spiral magnetic field in the inner 0.5~kpc points outward, which is {\em opposite}\ to that in the outer disk, and has a pitch angle of $\simeq33\grad$, which is much larger than that of $8\grad-19\grad$ in the outer disk. }
{The physical conditions in the central region differ significantly from those in the 10~kpc ring. In addition, the orientation of this region with respect to the outer disk is completely different. The opposite magnetic field directions suggest that the central region is decoupled from the outer disk, and we propose that an independent dynamo is active in the central region.}


\date{Received 9 Dec 2013 / Accepted 29 July 2014}
\maketitle

\section{Introduction}

With its distance of only 750~kpc\footnote{$752\pm27$~kpc from luminosity of Cepheids \citep{2012ApJ...745..156R} or $744\pm33$~kpc using eclipsing binaries \citep{2010A&A...509A..70V}}, the Andromeda galaxy (M31) is the nearest grand spiral galaxy and the largest extragalactic object on the sky after the Magellanic Clouds. Its proximity allows us to reach spatial resolutions in this galaxy of less than a kiloparsec with single-dish observations and of less than 100~pc with interferometer data.

The M31 galaxy has been a primary target for single-dish radio telescopes that are able to scan the full angular extent of its emission with high sensitivity. It is thus no surprise that M31 was one of the first external spiral galaxies from which polarized radio continuum emission was detected \citep{beck78}. The turbulent and ordered\footnote{Linearly polarized
radio continuum emission traces \emph{ordered} magnetic fields, which can be either \emph{regular} (or
\emph{large-scale}) magnetic fields (preserving their direction over large scales) or \emph{anisotropic turbulent} fields (with multiple field reversals within the telescope beam). To distinguish between these two components observationally, additional Faraday rotation data is needed.} magnetic field components are concentrated in a torus-like structure at about 10~kpc radius \citep{beck82}. This star-forming ``ring'' is a superposition of several tightly wound spiral arms with small pitch angles seen under a high inclination. The origin of the ring is still a matter of discussion. It may either result from an interaction with M32, one of M31's satellite galaxies \citep{gordon06,block06}, or from a major merger event between two large galaxies that possibly formed M31 \citep{hammer10,fouquet12}. Faraday rotation measures ($RM$) derived from radio polarization data at two wavelengths showed that the large-scale field of M31 is
regular; it preserves its direction around $360\grad$ in azimuth and across several kiloparsecs in radius
\citep{berkhuijsen03,fletcher04}.

By now, we know that all observed spiral (and even some irregular) galaxies exhibit ordered magnetic fields with an average field strength of $5\pm3$~$\mu$G, while the random field generally reaches around three times the strength of the ordered field \citep{fletcher10}. In galaxies where Faraday rotation data are available, a significant fraction of the ordered field is regular. The presence of these regular fields can be best explained by the mean-field $\alpha-\Omega$ dynamo theory \citep{beck96, moss98} that has recently received support from high-resolution MHD simulations \citep{gressel08,gent13}. The mean-field $\alpha-\Omega$ dynamo requires the presence of weak magnetic seed fields and an interplay of turbulence and shear to generate and maintain a regular field. Turbulence is provided by supernova explosions, and shear is a consequence of the differential rotation of the gas in the galactic disk. However, the generation of a regular field in a large galaxy takes billions of years and may not yet have been completed in present-day galaxies \citep{arshakian09}. In particular, large-scale field reversals, as relics of the initial seed field, may persist until the present time \citep{beck94,moss12}.

The detection of M31's regular magnetic field was a breakthrough for galactic dynamo theory, and it is still considered the prototype of a dynamo-generated field. The structure of the field of M31 is unusually simple -- an almost purely axisymmetric and plane-parallel field has not yet been found in any other spiral galaxy. According to dynamo models, the toroidal field in the disk should be accompanied by a poloidal field that extends away from the disk, as observed around many edge-on galaxies \citep{Krause09} and supported by the distribution of polarized emission in mildly inclined galaxies \citep{sings3}. The non-detection of a radio halo around M31 \citep{graeve81,berkhuijsen13,giess13} means that either the total field in the halo is weak or that very few cosmic-ray electrons in the GeV range (responsible for synchrotron emission in the GHz range) make it to the halo, for instance, due to cooling.

Radio emission from inside the 10~kpc ring of M31 is restricted to the central region with a radius of about 1.5~kpc, leaving a gap between these two regions. 
The average strength of the total magnetic field in rings around the centre increases from $15\pm3~\mu$G at
0.2--0.4~kpc radius to $19\pm3~\mu$G at 0.8--1.0~kpc radius, which is about 3--4~times larger than in the 10~kpc ring
\citep{hoernes_centre}.
$RM$ from three polarized background sources shining through the gap indicate that the large-scale regular field also exists inside the ring \citep{han}, but it is not illuminated by cosmic-ray electrons. At a frequency of 350~MHz, the gap is less pronounced in total power, but also clearly present in polarization \citep{giess13}. Formation of massive stars and hence sources of cosmic rays are limited to radii $\geq3$~kpc \citep[e.g.][]{kang}. Diffusion of cosmic-ray electrons across the ring is hampered by the regular field because the diffusion coefficient across the field lines is smaller than along the field \citep{2006MNRAS.373..643S, 2008ApJ...673..942Y, buffie}. The strength of the radio synchrotron emission from the central region raises the question of the origin of cosmic rays in that region. Particle acceleration by magnetic reconnection near the galaxy's centre \citep{lesch} or in shock fronts possibly seen in H$\alpha$ emission \citep{jacoby} are under discussion.

Thus far all polarization studies concentrated on the 10~kpc ring, while very little is known about the magnetic field in the central region. A spiral pattern with a large pitch angle was found in the southern half of the central region \citep{beck82,berkhuijsen03}, but the low spatial resolution did not allow to investigate its origin. A connection with the spiral field in the outer ring is hardly possible because the gas disk of the central region has a much smaller inclination than the ring plane \citep{ciardullo,melchior}. It is not known whether the same dynamo can operate in two differently oriented disks or whether two dynamos can operate independently. With our polarization observations and Faraday rotation data with high angular resolution and sensitivity, we can investigate this question.

\section{Observations}

The M31 galaxy has been observed with the Very Large Array (VLA) at 3.6~cm (8.46~GHz) on 23 December 2005 and at 6.2~cm (4.86~GHz) on 26 July 1992. Both observations were made using the D-Array. The 6.2~cm data has already been presented in \cite{hoernesdr} but was reduced again from the raw UV data for this analysis, since the images were contaminated by side-lobes from a bright off-centre source (see below).

The Effelsberg data was taken over several years (1999 (8.35~GHz, 3.6~cm); 2001--2005 (4.85~GHz, 6.2~cm)) and will be presented in a separate paper (Gie\ss\"ubel et al. in prep.). The 3.6~cm receiver is a single-horn receiver with a bandwidth of 1.1~GHz centred on 8.35~GHz; the 6~cm receiver is a dual-horn receiver with a bandwidth of 0.5~GHz centred on 4.85~GHz. Both are installed in the secondary focus of the Effelsberg 100-m Telescope. Both maps were observed using on-the-fly mapping, alternating in two perpendicular directions.

The observational details are summarized in Table~\ref{tab:obsum}.

\begin{table}
\caption{Observational details.}
\label{tab:obsum}
\centering
\begin{tabular}{ll}\hline\hline
Centre position & \textsc{Ra} = $00^\text{h}\,42^\text{m}\,46^\text{s}.0485$\\
 (J2000) & \textsc{Dec} = $41\grad\,16\arcmin\,11\arcsec.623$ \\
Distance & $750\pm30$~kpc\\
Angular resolution & 15\arcsec \\
Along major axis & 15\arcsec $\hatap$ $54.5\pm2.2$~pc\\
Rms noise$^a$\\
3.6~cm total power & $\sigma_t=5.5$~$\mu$Jy/beam\\
3.6~cm $PI$ & $\sigma_{PI}=4.8$~$\mu$Jy/beam\\
6.2~cm total power & $\sigma_t=7.5$~$\mu$Jy/beam\\
6.2~cm $PI$ & $\sigma_{PI}=6.8$~$\mu$Jy/beam\\\hline
\multicolumn{2}{l}{$^a$ The noise values refer to the central area of the primary}\\
\multicolumn{2}{l}{ beam; they increase with distance from the field centre.}
\end{tabular}
\end{table}

\section{Data reduction}
Data reduction of the VLA data was made in \texttt{AIPS}\footnote{http://www.aips.nrao.edu/} using the standard tasks and \texttt{VLACALIB}. At both wavelengths, 3C48 was used as a flux calibrator and 3C138 for polarization calibration (polarization angle $-12\grad$). As phase calibrators, 3C19 was observed at 6.2~cm and 0013+408 (J2000) at 3.6~cm.

North of the centre, a very bright background source, 37W115 (\citealt{walterbos}), is located that is also bright in polarization (e.g. \citealt{beck89}). This source had to be removed from the 6.2~cm data. The \texttt{AIPS} task \texttt{PEELR} was used to subtract the source from the uv data.
At 3.6~cm the primary beam is much smaller and no obvious side-lobes of 37W115 were disturbing the image, so that no peeling was needed for this dataset.

\begin{figure*}[ht]
\centering
\subfigure[3.6~cm total power map]{
    \includegraphics[width=0.36\textwidth,angle=-90]{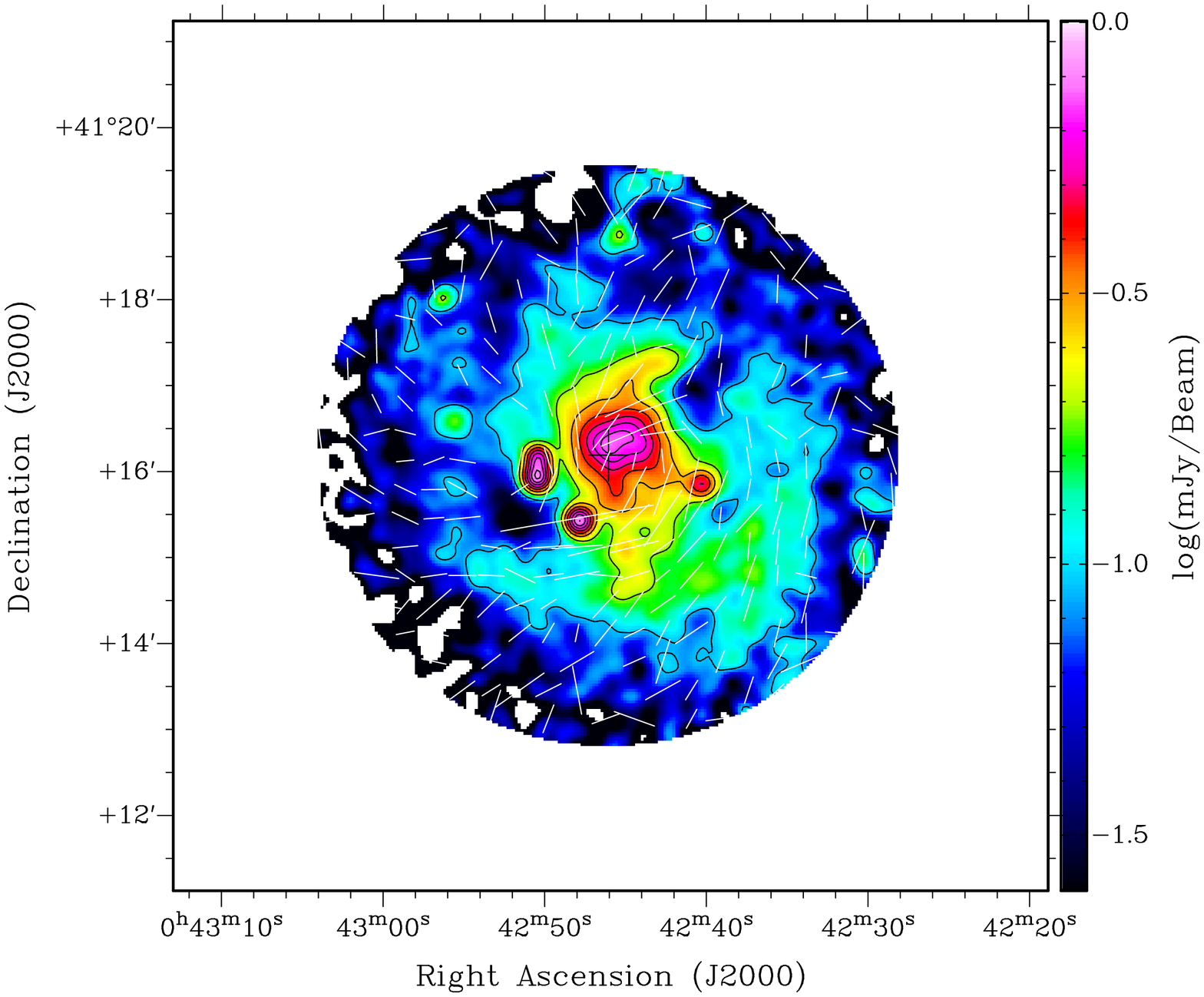}
    \label{3cmvla}
}
\subfigure[3.6~cm polarized intensity map]{
    \includegraphics[width=0.36\textwidth,angle=-90]{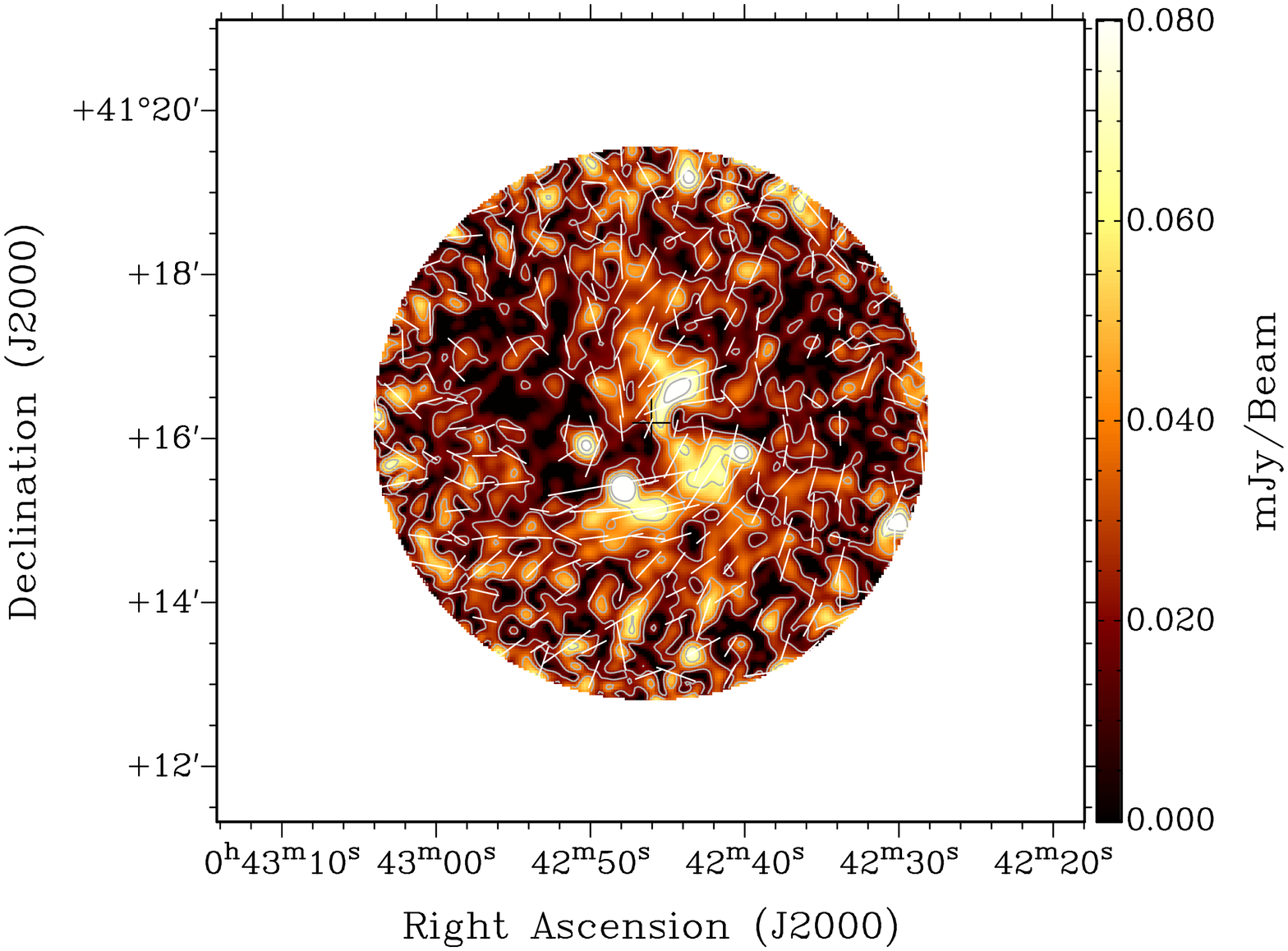}
    \label{3cmvlaPI}
}
\subfigure[6.2~cm total power map]{
    \includegraphics[width=0.36\textwidth,angle=-90]{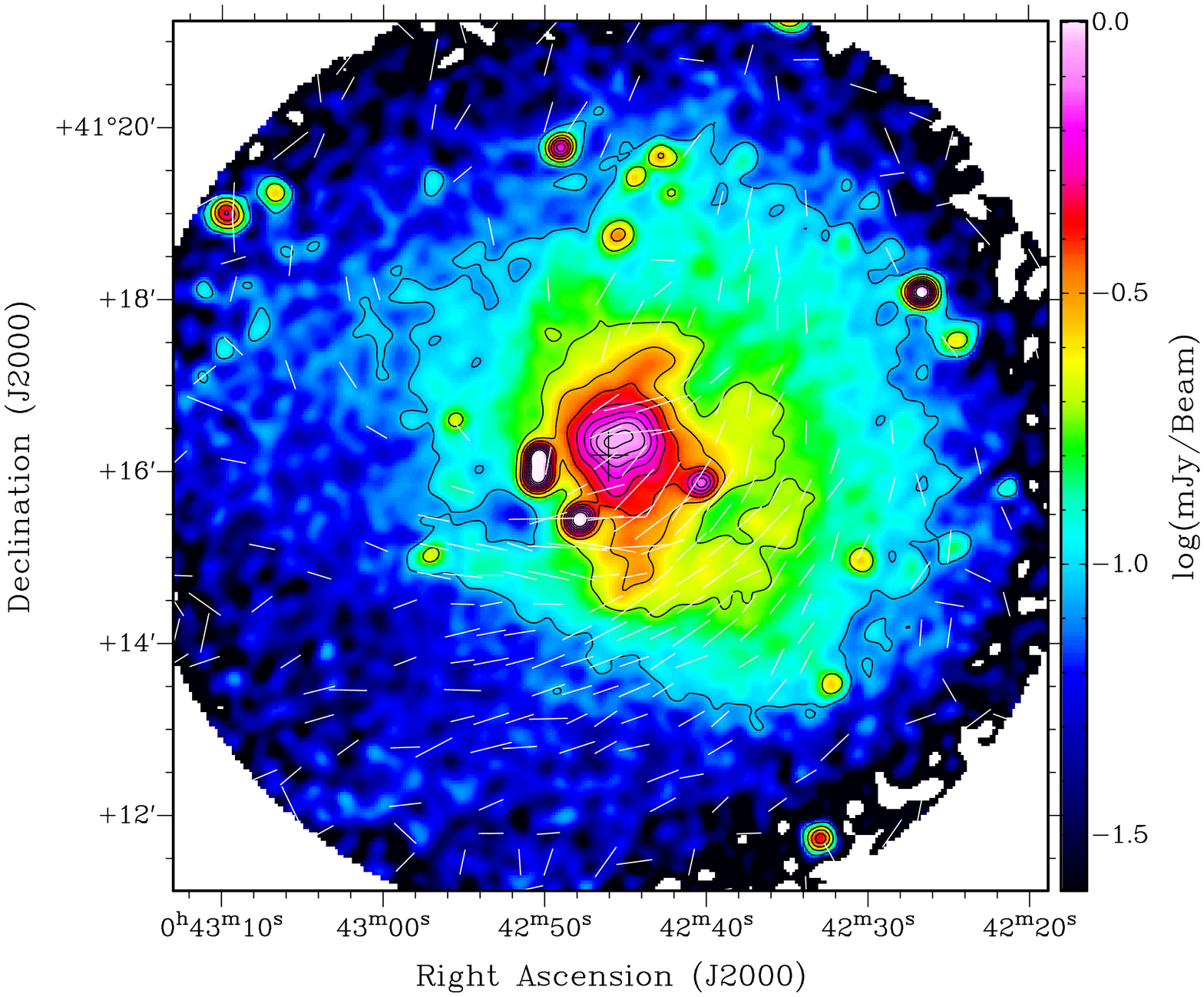}
    \label{6cmvla}
}
\subfigure[6.2~cm polarized intensity map]{
    \includegraphics[width=0.36\textwidth,angle=-90]{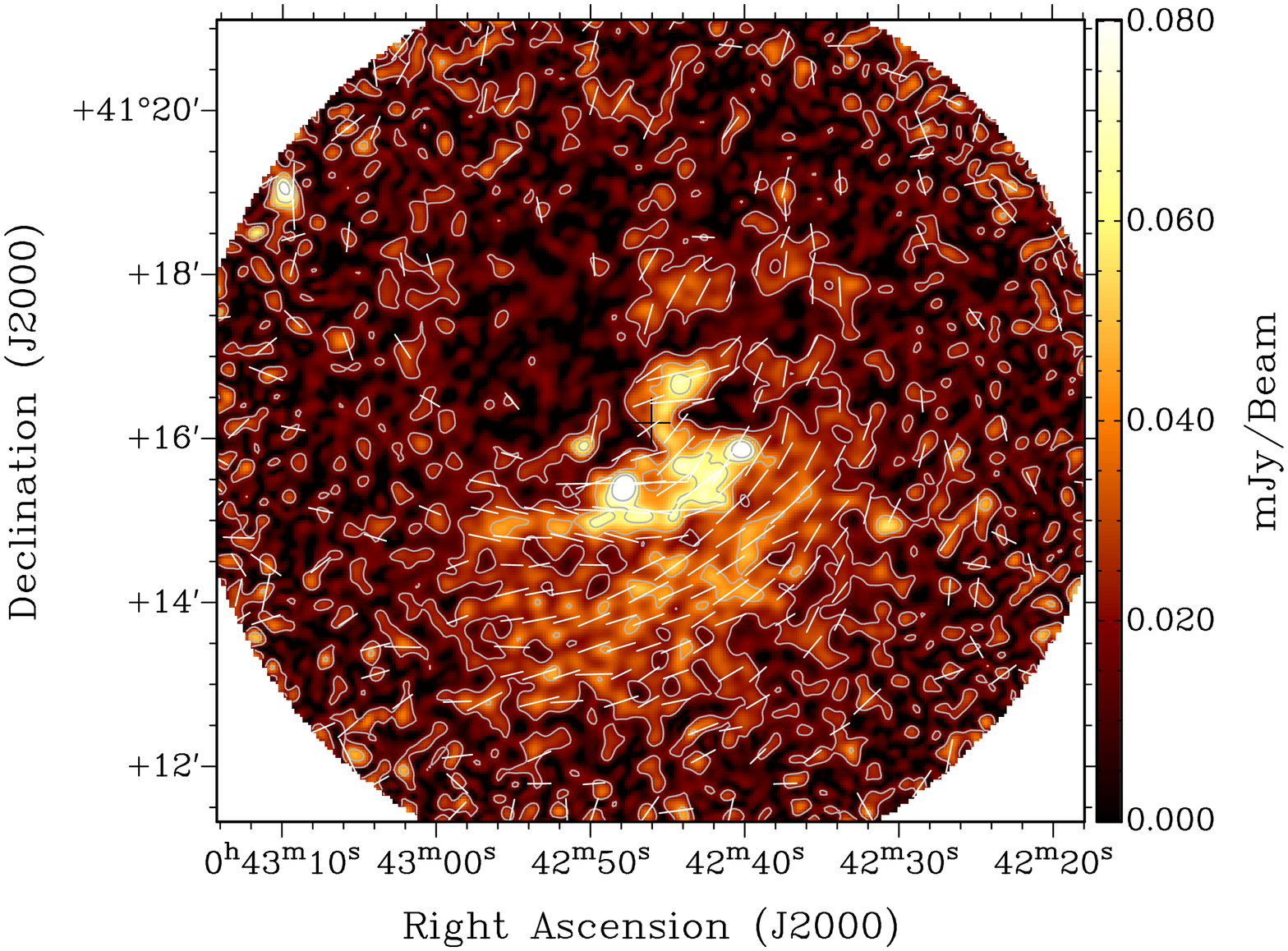}
    \label{6cmvlaPI}
}
\caption[Total power and polarized intensity maps of the combined VLA and Effelsberg data.]{Total power and polarized intensity maps of the combined VLA and Effelsberg data at $\lambda$3.6~cm and $\lambda$6.2~cm at 15\arcsec\ resolution. The pointing centre of the VLA maps is marked with a black cross. Vectors show the orientation of $\vec{E}+90\grad$, which are not corrected for Faraday rotation of the foreground. A vector-length of 1\arcsec\ corresponds to a polarized intensity of 0.01 mJy, they are cut off at $3\sigma_{PI}$. The range of $RM$ between about -200 rad/m$^2$ and +100 rad/m$^2$ (see Fig.~\ref{fig:model-on-data}) corresponds to rotation angles $-15\grad$ to $+7.5\grad$ at 3.6~cm and $-44\grad$ to $+22\grad$ at 6.2~cm. The depicted map size is $\sim9'\times9'$. Contours of the total power maps range from 90 to 890 $\mu$Jy/beam in steps of 100~$\mu$Jy/beam on a logarithmic colour-scale. Contours of the polarized intensity maps range from 20 to 100~$\mu$Jy/beam in steps of 20 $\mu$Jy/beam on a linear colour scale.}
\label{vla_imgs}
\end{figure*}

\subsection{Combination with Effelsberg data}\label{vlaeffcomb}

Interferometric data from a synthesis telescope usually suffer from missing zero spacings: The uv plane is never filled entirely, and there is especially a gap between the shortest baselines of the array and zero baseline. The consequence is that a synthesis telescope misses extended emission; it is only sensitive to structures up to $\lambda / D_{min}$, where $D_{min}$ is the shortest baseline. The gap in the spatial-frequency domain results in a depression of the response function. In the resulting images this becomes apparent by sources seemingly ``sitting in a bowl of negative emission''.

The gap in the uv range can be filled using an observation at the same frequency taken with a single-dish telescope that is larger than the shortest spacing of the synthesis telescope. This process is also referred to as {\em feathering}. To merge the Effelsberg and VLA data, the \texttt{AIPS}-task \texttt{IMERG} was used. It combines two images in Fourier space.
The merged maps are shown in Figures~\ref{3cmvla} to~\ref{6cmvlaPI}. The maps have a resolution of $15''$. Please refer to the image captions and Table~\ref{tab:obsum} for further details.

The central region is seen from below. The northern side is thus closer to the observer than the southern side. Also, the north-eastern side is the receding and the south-western the approaching side \citep[e.g.][]{boulesteix}.

The total emission is brightest in the centre. Additionally the southern arm known from H$\alpha$ emission is clearly visible. Note, however, that the polarized emission is strongest on the inside of the H$\alpha$ arm and vanishes at the actual position of the arm. Especially at 6.2~cm, the southern arm is significantly depolarized, probably due to Faraday depolarization. The enhancement of emission on the inner side of the arm indicates compression of the magnetic field lines due to weak shocks.

Furthermore, a large part of the emission at 6.2~cm in the north-east is completely depolarized. The implications of this polarization pattern are discussed in Section~\ref{sect:3dstruct}.

\section{Spectral index}

\begin{figure}
\begin{center}
\includegraphics[scale=0.3,angle=-90]{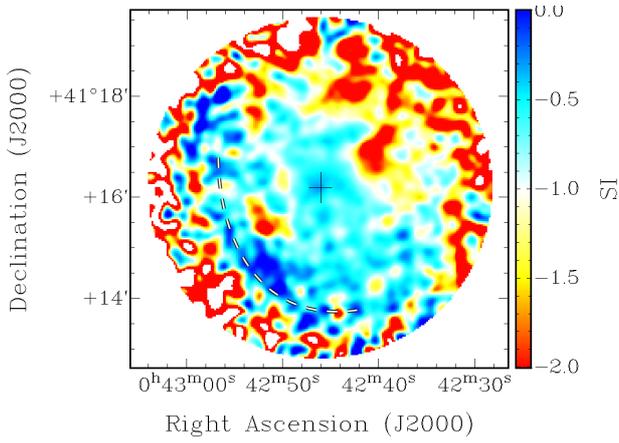}
\caption{Spectral index map between 3.6~cm and 6.2~cm at 15\arcsec\ resolution clipped above 0 and below -2. The dashed line indicates the approximate position of the H$\alpha$ arm \citep{jacoby}.}
\label{fig:SImap}
\end{center}
\end{figure}

Figure~\ref{fig:SImap} shows the spectral index ($SI$) map between 3.6~cm and 6.2~cm. Typical errors for SI are $\pm0.02$ in the centre, $\pm0.2$ on the southern arm, and $\pm2$ towards the edge of the map. There are two features, which can be clearly distinguished in the $SI$ map: the inner 400 pc and the southern arm, known from H$\alpha$ observations \citep[e.g.][]{jacoby}. The spectral index in the very centre is $-0.40\pm0.03$ and steepens outwards where it quickly reaches values of $-1.0$ and steeper. A similar behaviour was already noted by \cite{hoernes_centre} between 6.2~cm and 20~cm.
There is little H$\alpha$ emission \citep{jacoby} and hence little star formation in the centre, so that the flat spectrum is most probably due to synchrotron emission. \cite{hoernes_centre}
proposed that the cosmic ray electrons in the inner kiloparsec originate in a mono-energetic source in the centre, which is associated with the central black hole, as proposed for the centre of the Milky Way \citep{lesch88}. However,
such a source should generate an inverted synchrotron spectrum, which is not the case in M31. We propose that the
cosmic-ray sources are related to nuclear activity, such as shocks in gas outflows or inflows, or stellar winds from the
cluster of blue stars, as reported by \cite{lauer}.

The spectral index of the southern arm is again flat with average values between $0.0$ and $-0.5$ ($\pm0.1$). In the region where it flattens to 0 (\textsc{Ra}=$00^\text{h}\,42^\text{m}\,52^\text{s}$ \textsc{Dec}=$+41\grad\,14\arcmin.5$), the polarized intensity is reduced by 10 to 20~$\mu$Jy/beam on average. Since this position coincides with the H$\alpha$ arm, we conclude that the emission is dominated by thermal emission and the synchrotron contribution
is small in this region. If a significant number of cosmic ray electrons is accelerated by strong shocks or supernova remnants
in the southern arm, we would expect a radio spectral index similar to that in normal spiral arms, which around $-0.7$.
Therefore, the very centre is a more likely source of cosmic ray electrons that we can identify.

\section{Sector analysis of the magnetic field structure}\label{sec:sector}

The inclination $i$ of the plane of the central region and the position angle $pa$ of its major axis differ significantly from the inclination and position angle of the outer disk. While these parameters are very well defined for the 10~kpc ring ($i=75\grad$, $pa=38\grad$, e.g. \citealt{chemin}), literature values for the inner kiloparsec vary a lot. Usually an inclination of $i=45\grad$ \citep{ciardullo} is assumed for the nuclear spiral, but the position angle is not well defined due to this mild inclination. Values range from $pa\approx38\grad$ like the major axis of the 10~kpc-ring \citep[e.g.][]{boulesteix,ciardullo} to $pa=48\grad$ \citep{saglia} and $i=43\grad$, $pa=70\grad$ \citep{melchior}.

For the following analysis we use $i=43\grad$ and $pa=70\grad$, which is in best accordance with our results. We are, however, not able to further constrain the position angle. We show plots for the other position angles in Section~\ref{sec:robustness}, where we also address the robustness of our findings with regards to different position angles.

\begin{figure}
\begin{center}
\includegraphics[scale=0.5]{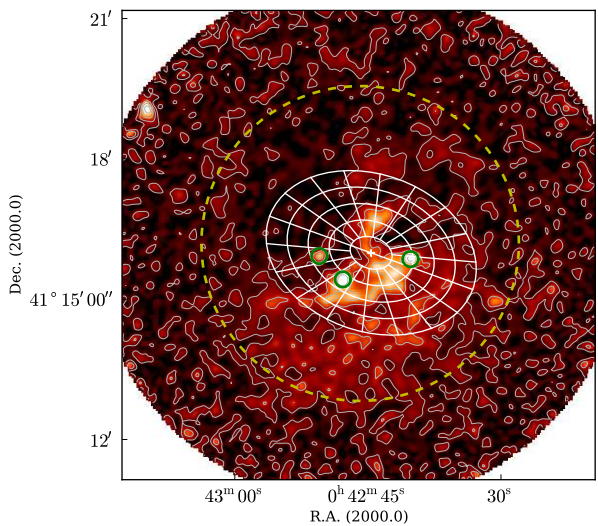}
\caption{Sectors for $i=43\grad$ and $pa=70\grad$ on the 6.2~cm polarized intensity map. Rings start at $r=0.1$~kpc with $\Delta r=0.1$~kpc, the azimuthal width of the sectors is $20\grad$. They are counted counter-clockwise, starting at the northern side of the major axis. The dashed yellow line shows the size of the primary beam of the 3.6~cm map. The three strong polarized point sources marked with green circles have been subtracted from the Q and U images. Contours are the same as Fig.~\ref{6cmvlaPI}.}
\label{fig:sector}
\end{center}
\end{figure}

We define four rings with a radial width of 0.1~kpc in the plane of the central disk, starting at $r=0.1$~kpc (assuming a distance of 750~kpc to M31). Each ring is divided into sectors with an azimuthal width of $20\grad$ in the disk plane. Figure~\ref{fig:sector} shows these sectors on top of the 6.2~cm $PI$ map. The azimuthal angle is counted counter-clockwise, starting at the northern side of the major axis. The central position of the sectors is slightly offset from the pointing centre at \textsc{Ra}=$00^\text{h}\,42^\text{m}\,44^\text{s}.45$ \textsc{Dec} =$41\grad\,16\arcmin\,1\arcsec.62$ (J2000), so that the sectors are centred on the spiral pattern visible in polarized intensity.

For each sector, we computed the average polarized intensity $PI$ at 6.2~cm, the average $RM$ between 3.6~cm and 6.2~cm, and the pitch angle $\phi_{\rm obs}$ of the projected average $\vec{B}$ vectors against the ring orientation as follows:

\begin{equation}
PI = \frac{1}{N}\sum^N_i \sqrt{Q^2_i + U^2_i - (1.2\sigma)^2},
\end{equation}
where $Q_i$ and $U_i$ is the value at the $i^{\rm th}$ pixel in the 6.2~cm Stokes $Q$ and Stokes $U$ map, respectively; $\sigma$ is the average noise of the $Q$ and Stokes $U$ maps and $N$ the number of pixels in the sector.

\begin{equation}
RM = \frac{\Delta \mathcal{X}^2}{\Delta \lambda^2},
\end{equation}
where $\Delta \mathcal{X}$ is the difference between the sector averages of the polarization angles at 6.2~cm and 3.6~cm calculated as

\begin{equation}
\mathcal{X}_\lambda = \frac{1}{2} \arctan \frac{\langle U_\lambda \rangle}{\langle Q_\lambda \rangle},
\end{equation}
here $\langle U_\lambda \rangle$ and $\langle Q_\lambda \rangle$ denote the sector averages of the Stokes $Q$ and Stokes $U$ map at wavelength $\lambda$. The pitch angle is

\begin{equation}
\phi = - \theta \pm 90\grad + \arctan\left(\frac{\tan(\mathcal{X}_0-pa)}{\cos i}\right),
\end{equation}
where $\theta$ is the azimuthal angle of the sector and $\mathcal{X}_0 = \mathcal{X}_\lambda - RM \lambda^2 \pm 90\grad$, the intrinsic position angle of the $B$-vector in the plane of the sky. Following the general convention, a positive pitch angle indicates a clockwise winding, and negative pitch angles indicates a counter-clockwise winding spiral. We note that this definition is opposite to that used by \cite{fletcher04}; we, thus, changed the sign of the cited values according to the definition in this paper.

For a planar magnetic spiral field, the pitch angle represents the pitch angle of the spiral in the plane of the central region. The three strong polarized background sources (marked green in Fig.~\ref{fig:sector}) have been subtracted from the Q and U images before the analysis.

\begin{table}
\caption{Average pitch angle of the spiral field in the four rings.}
\label{tab:pitch}
\centering
\begin{tabular}{cc}\hline\hline
mean radius [kpc] & $\phi$ [$\grad$]\\
0.15 & $30\pm5$\\
0.25 & $33\pm3$\\
0.35 & $33\pm3$\\
0.45 & $33\pm3$\\\hline
\end{tabular}
\end{table}

Table~\ref{tab:pitch} lists the weighted average pitch angle of the spiral field in each of the four rings. Although the observed pitch angle varies significantly over the azimuthal range, the average still represents the pitch angle of the spiral field component (see Sect.~\ref{sec:robustness}).

The sector averages for all rings are shown in Fig.~\ref{fig:model-on-data}. The sector averages of $PI$ clearly show the depolarization of the emission at the northern part of the major axis at 6.2~cm. The sector averages of $RM$ show the characteristic pattern expected for an axis-symmetric spiral (ASS) magnetic field in the disk \citep[e.g.][]{krause89}. As mentioned above, the pitch angle varies significantly in a coherent pattern, which is is not expected for a magnetic field confined to a disk.

\section{Large-scale magnetic field reversal in M31}\label{sect:reversal}

The direction of the regular magnetic field is non-ambiguously determined by the observed $RM$ pattern\footnote{We emphasize that the result is purely observational and that no assumptions about the overall field geometry are required.}. The entire pattern is shifted by the foreground $RM$ of -93~rad~m$^{-2}$ \citep{han,berkhuijsen03,fletcher04}, as indicated by the red dashed line in Fig.~\ref{fig:model-on-data}, middle column. If the intrinsic $RM$ of the source is positive, the magnetic field vectors are pointing away from the observer; if the intrinsic $RM$ of the source is negative, the magnetic field vectors are pointing towards the observer. Following \cite{krause89}, we find that the spiral field in the inner 0.5~kpc is pointing outwards, as depicted in Figure~\ref{fig:model}.

\cite{fletcher04} studied the regular magnetic field of M31 using observations at 6~cm, 11~cm, and 20~cm. They found that the magnetic field between 6~kpc and 14~kpc is pointing inwards with pitch angles ranging from $\phi=19\grad$ to $\phi=8\grad$, which become smaller with increasing radius. The large-scale fields in the central region and in the 10~kpc ring are thus {\em pointing into opposite directions}. The reversal itself happens in the radial range between 1.5~kpc and 6~kpc, where the strength of the regular field has a minimum \citep{beck82,moss98}.

Radial field reversals can be explained by the mean-field $\alpha-\Omega$ dynamo theory as relics of the initial seed field \citep{beck94,moss12,moss+sokoloff12}. Fields of opposite polarity are stretched by differential rotation and eventually reconnect. One polarity wins, leaving a coherent axisymmetric disk field as the final configuration \citep{hanasz}. (We note that the sense of the winning polarity is random; it is neither determined by the initial conditions nor by the sense of rotation of the galaxy.) Tidal interactions or continuous injection of turbulent fields may distort or slow down this process \citep{arshakian09,moss12}. \cite{ferriere+schmitt} showed that a radial field reversal can also be explained in the framework of the $\alpha-\Omega$ dynamo theory by a vertical gradient in the galactic rotation rate. Such a gradient is most likely present in the central disk of M31 due to the ever-changing inclination and position angles over the entire disk \citep[e.g.][]{chemin}.

As the central region of M31 is decoupled from the 10~kpc ring due to its entirely different inclination and position angle, reconnection of fields of the inner disk and the ring with different polarities seems impossible. Furthermore, the physical conditions in the central region are different from those outside (i.e. little star formation in the central region but a star-forming ring in the outer part, which coincides with the 10~kpc ring seen in radio emission). For these reasons, we propose that the magnetic fields in the inner ($\lsim4$~kpc) and outer part ($\gsim4$~kpc) of M31 are generated independently, as was already suggested by \cite{ruzmaikin81}. As the field polarity is a random number, the reconnection process of seed fields has developed differently in the two dynamo-active regions.

The efficiency $\eta$ (given by the dynamo number) and timescale $t_\mathrm{dyn}$ of the mean-field
$\alpha-\Omega$ dynamo in the thin-disk approximation can be estimated by \citep{ruzmaikin88}:
\begin{equation}
\eta \simeq 9\, h^2 \, \Omega \, |S| \, / v_\mathrm{turb}^2\, ,
\end{equation}
\begin{equation}
t_\mathrm{dyn} \simeq 3 \, h \, / ( (\Omega \, |S| )^{0.5} \, l_\mathrm{turb}).
\end{equation}
For a flat rotation curve ($v_\mathrm{rot}$=const):
\begin{equation}
\eta \simeq 9\, (h/R)^2 \,\, (v_\mathrm{rot}/v_\mathrm{turb})^2\, ,\label{eta}
\end{equation}
\begin{equation}
t_\mathrm{dyn} \simeq 3 \, h \, R / (v_\mathrm{rot} \, l_\mathrm{turb})\, ,\label{time}
\end{equation}
where $\Omega$ is the angular rotational velocity, $S = R (\partial \Omega/\partial R)$ is the rotational shear,
$v_\mathrm{turb}$ is the turbulent velocity, $l_\mathrm{turb}$ is the basic length scale of turbulence, and $h$
is the scale height of the ionized gas. The turbulent velocity $v_\mathrm{turb}$ is known to be mostly independent of the star formation rate (except for starbursts) \citep{2006ApJ...638..797D}, so $v_\mathrm{turb}$ can be assumed to be similar in the inner and outer disks of M31, and $S$ and $h$ are the main driving parameters. The scale height $h$ is probably about three times smaller in the inner disk than that in the outer disk (see below). The rotation curve is roughly flat in the inner disk beyond $1\arcmin \simeq 0.2$~kpc \citep{boulesteix}, as well as in in the outer disk \citep{corbelli}, and the rotation velocities are 200--250~km/s. 
According to Eqs.~\ref{eta} and \ref{time}, the dynamo efficiency scales with $(h/R)^2$ and the dynamo time-scale with $h \cdot R$. With a radius ratio of about 1/10 and a height ratio of about 1/3, the dynamo in the inner disk is about $10\times$ more efficient and about $30\times$ faster than that in the outer disk.

In the thin-disk approximation $\alpha-\Omega$ dynamo, the pitch angle of the spiral field can be estimated as $\phi \approx \arctan(l_\mathrm{turb}/h)$ \citep{beck96}. Little is known about the variation of $l_\mathrm{turb}$ within galaxies, so that we assume $l_\mathrm{turb}$=const here. The observed ratio of pitch angles in the inner and outer disks is about 3, and by using $h\approx$1~kpc in the outer disk \citep{fletcher04}, this indicates $h\approx$370~pc in the inner disk. With a radius of the inner disk of about 1.5~kpc, a scale height of about 370~pc is still consistent with a thin disk.

\section{3-D structure of the central magnetic field}\label{sect:3dstruct}

\subsection{Comparison with the Westerbork SINGS survey}
The strong variations of the observed pitch angle with the azimuthal angle suggest that a 2-D configuration is not sufficient to explain the observed sector averages (Fig.~\ref{fig:model-on-data}). Furthermore, the pattern of the observed polarized emission at 6.2~cm (Fig.~\ref{6cmvlaPI}) shows a striking resemblance to the polarized intensity images of the spiral galaxies observed in the Westerbork SINGS survey, in which a sample of large, northern galaxies from the Spitzer Infrared Nearby Galaxies Survey (SINGS) were observed with the WSRT in two bands at 1300--1432~MHz and 1631--1763~MHz \citep{sings1}: For all moderately inclined galaxies with extended polarized emission, the polarized flux is lowest on the side of the major axis of the projected galaxy disk that is receding and highest on the approaching side \citep{sings2,sings3}.

\begin{figure*}
\centering
\subfigure[In-plane component of the vector field seen face-on. The spiral pitch angle is $\phi=33\grad$.]{
    \includegraphics[trim=100px 40px 100px 40px,clip,width=0.45\textwidth]{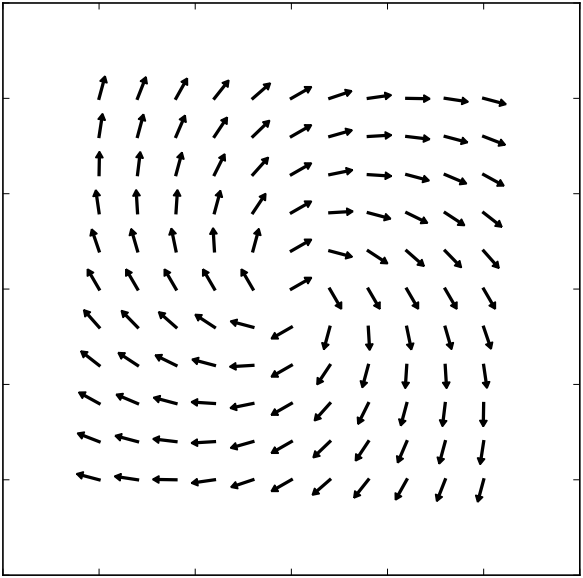}
    \label{model_faceon}
}\hfill
\subfigure[Out-of-plane ``divergence-free'' component of the vector field seen edge-on. We observe the field at an inclination. The far side is depolarized at 6.2~cm, so that only the near side is visible to us (indicated by the grey-scale).]{
     \includegraphics[trim=200px 80px 200px 80px,clip,width=0.45\textwidth]{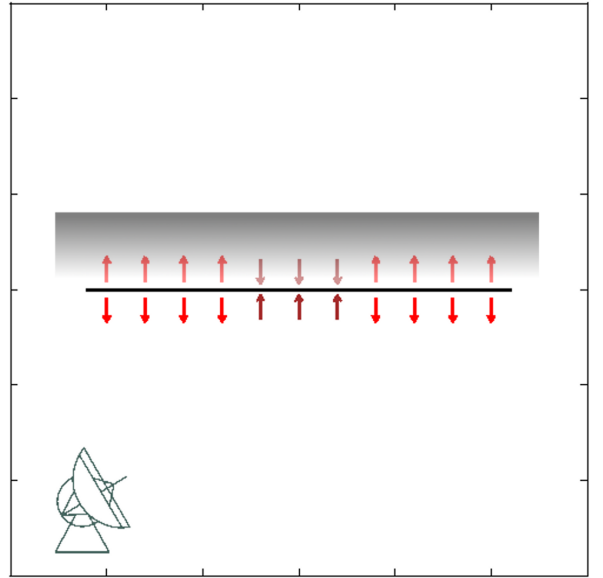}
     \label{model_edgeon}
}
\subfigure[Spiral (black) and vertical (red/brown) components of the field in the sky plane seen under the inclination $i=43\grad$ and the position angle $pa=70\grad$. Only the near side is visible to us.  The kinematically receding side is to the northeast.]{
    \includegraphics[trim=100px 40px 100px 40px,clip,width=0.45\textwidth]{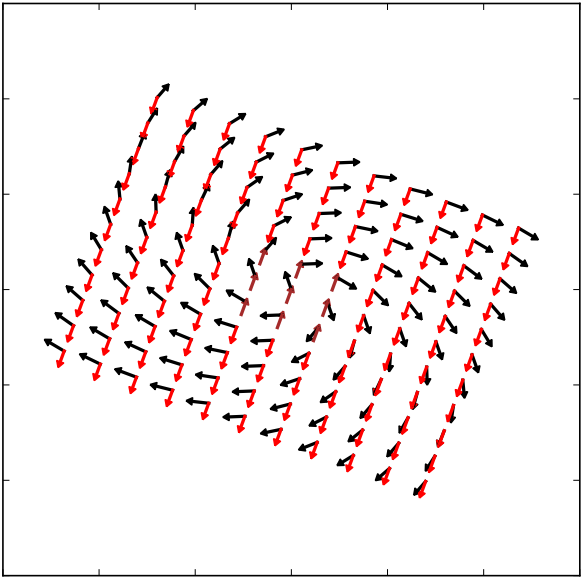}
    \label{model_incl}
}\hfill
\subfigure[The full vector field of the near side, as seen at 6.2~cm under the inclination $i=43\grad$ and the position angle $pa=70\grad$. Contours of Fig.~\ref{6cmvlaPI} are underlaid for comparison. For better perceptibility, the lowest contour is filled yellow.]{
    \includegraphics[trim=100px 40px 100px 40px,clip,width=0.45\textwidth]{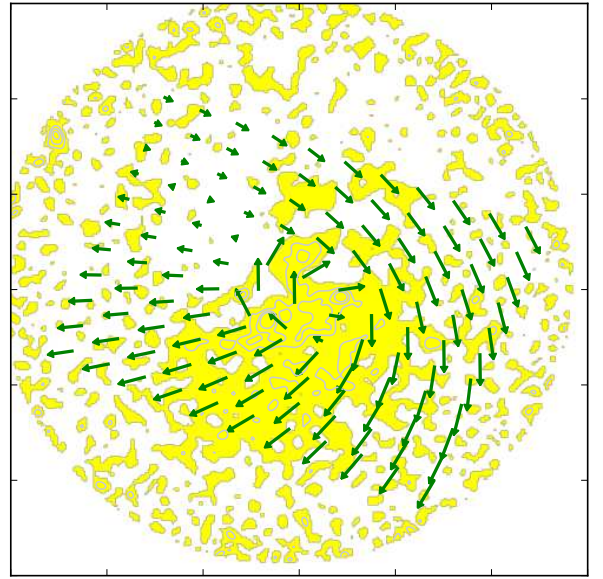}
    \label{model_incl_add}
}
\caption[]{Proposed field geometry for the inner 0.7~kpc of the central disk, as seen in projection under different inclinations. The black vectors depict the component in the xy-plane of the disk; the red vectors show the component in the z-direction vertical to the disk. The green vectors show the vector addition of both components.}
\label{fig:model}
\end{figure*}

\begin{figure*}[htbp]
\begin{center}
\includegraphics[scale=0.55, bb=-77 -102 674 884]{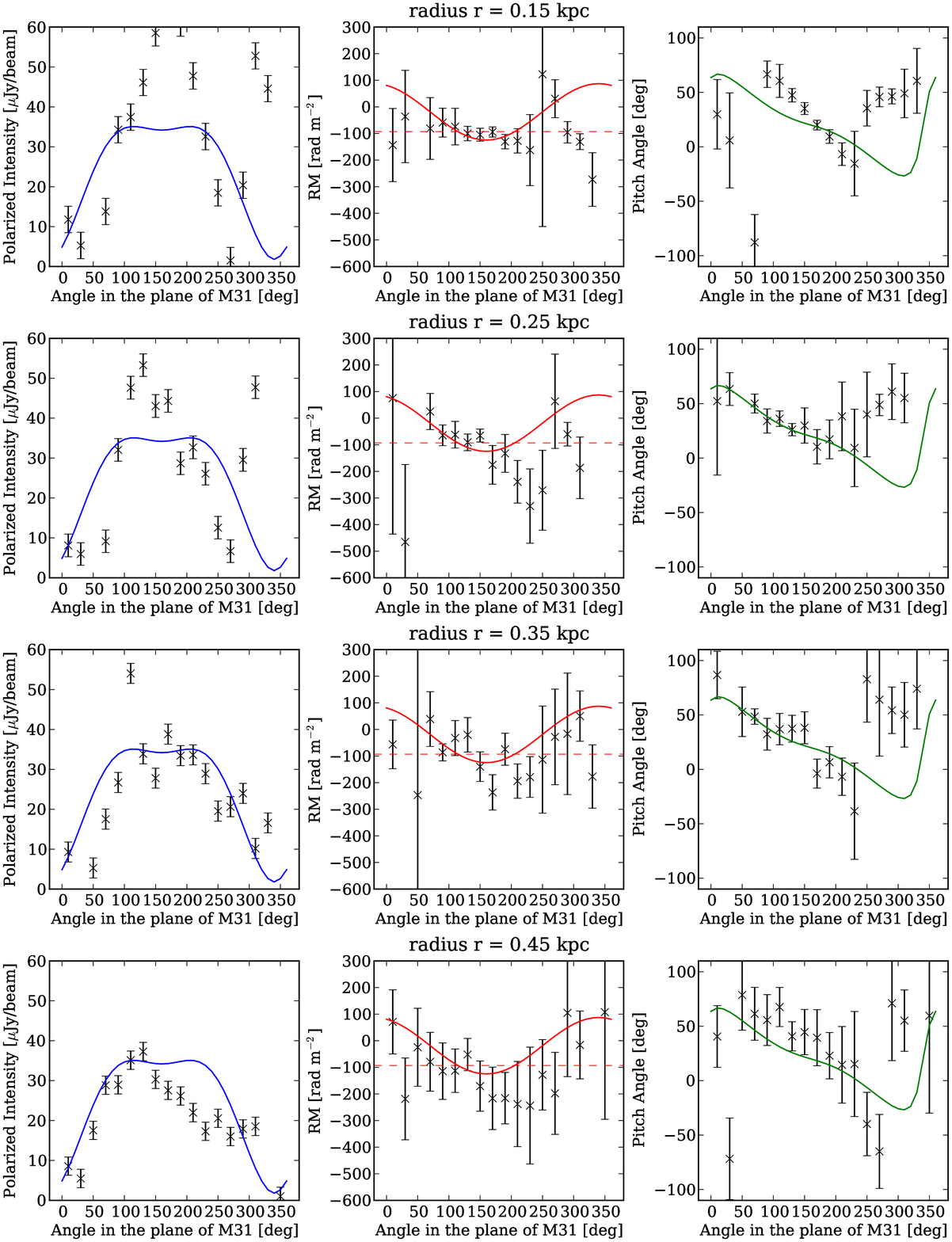}
\caption{The points with error bars show (from left to right) the observed values of $PI$ at 6.2~cm, $RM$ between 3.6~cm and 6.2~cm and $\phi_{\rm obs}$ respectively, for the sectors described in Section~\ref{sec:sector}. From top to bottom, the results for the rings with mean radius 0.15~kpc, 0.25~kpc, 0.35~kpc and 0.45~kpc are presented. The lines are no fits but show the model-output (see Sect.~\ref{sect:model}) using $B_{\rm sp}=1.0$, $B_{\rm ver}=0.7$, $\phi=30.0\grad$, $c_{pi}=20.0$, $c_{rm}=150.0$, $i=43\grad$, and $pa=70\grad$ for all rings. The red dashed line shows $RM_{\rm fg}$; the expected rotation measure of the Galactic foreground, by which the model was shifted.}
\label{fig:model-on-data}
\end{center}
\end{figure*}

In the case of the central region of M31, it is also the receding side that is depolarized (compare Fig.~\ref{6cmvlaPI} with the velocity field of the central region shown in Fig.~5 of \cite{melchior}).
As there is no indication for enhanced turbulent fields or an enhanced ionized gas density (H$\alpha$ emission)
on the receding side compared to the approaching side, we propose that the observed pattern in polarized
intensity and the asymmetry with respect to the major axis are the result of the 3-D field topology projected
onto the sky plane and of Faraday effects, as proposed by \cite{sings3}. 
They assume that (1) the field structure is of
even symmetry (of quadrupolar type), (2) the spiral pattern in the disk follows the optical spiral arms that
are trailing with respect to the galaxy rotation, and (3) polarized emission from the far side of the halo
is depolarized by Faraday dispersion, so that only polarized emission from the disk and the front
side of the halo facing the observer remains visible. Then, in an inclined galaxy disk, the components of the
magnetic field of the disk and halo in the sky plane add on the approaching side but partly cancel on the receding
side of the major axis.

In the disks of spiral galaxies, sufficiently strong depolarization that fulfils assumption (3) generally occurs at
wavelengths of $\ge$20~cm. In the central region of M31, however, strong depolarization on the receding side occurs
already at 6.2~cm. The polarization degree $p=PI/I$ on the receding side is $p_6\simeq6$~\% at 6.2~cm
and $p_{3.6}\simeq20$~\% at 3.6~cm, while the numbers are similar
($p_6\simeq17$~\% and $p_{3.6}\simeq20$~\%) on the approaching side. As the polarized intensity on the receding side
is {\em not}\ reduced at 3.6~cm, the far side of the halo is not depolarized at this wavelength.

Depolarization $DP$ by Faraday dispersion can be estimated by

\begin{equation}
DP = \frac{1-exp(-2 \sigma_{RM}^2 \lambda^4)}{2 \sigma_{RM}^2 \lambda^4},
\end{equation}
where $\sigma_{RM}$ is the RM dispersion \citep{sokoloff98,giess13} that can be expressed as
$\sigma_\mathrm{RM}\,=\,0.81\, <n_\mathrm{e}>\, B_\mathrm{r}\, d\, (L/(d\,f))^{0.5}$ \citep{beck07},
where $<n_\mathrm{e}>$ is the average thermal electron density of the diffuse ionized gas along the line
of sight (in cm$^{-3}$), $B_\mathrm{r}$ the random field strength (in $\mu$G), $L$ the pathlength through the
thermal gas (in pc), $d$ the turbulent scale (in pc) and $f$ the volume filling factor of the Faraday-rotating gas.
Significant depolarization (say, $DP\simeq0.6$) at 6.2~cm needs $\sigma_\mathrm{RM}\simeq200$~rad\,m$^{-2}$.
This can be achieved by $<n_\mathrm{e}> \simeq0.055$~cm$^{-3}$, $B_\mathrm{r}\simeq20~\mu$G \citep{hoernes_centre},
$L=370/cos(i)\simeq500$~pc (Sect.~\ref{sect:reversal}), $d\simeq50$~pc and $f\simeq0.5$.

\subsection{Our magnetic field model}\label{sect:model}

To explain the observed polarization pattern, we use a simplified geometrical model, similar to that of \cite{sings3}, to calculate the magnetic field components along and perpendicular to the line of sight and the apparent pitch angle in the sky plane. The vertical field component has the pattern of a quadrupolar or dipolar\footnote{With our simple model we can actually not distinguish between a dipolar or quadrupolar-type field (see below). According to the mean-field $\alpha-\Omega$ dynamo model, a quadrupolar-type field is preferred in a thin galactic disk.} and the horizontal field that of an axisymmetric spiral. Following our observation, the radial field component points outwards (Sect.~\ref{sect:reversal}). As a consequence, the vertical component of the poloidal field points towards the plane in the inner region and away from the plane in the outer region. We can decompose this pattern into a log-spiral field in the plane of the disk of the galaxy and a perpendicular component in the vertical direction.

We define the vectors in a fixed coordinate system for a galaxy seen face-on and then rotate the vectors accordingly. We use a right-handed, three-dimensional, Cartesian coordinate system, where the positive y-axis points towards the observer. The plane of a galaxy seen face-on thus resides in the xz-plane with the positive z-axis pointing to the north and the positive x-axis pointing to the east.

For any position $\vec{r}$ in the galaxy at distance $r$ from the centre and at an azimuthal angle $\theta$ counted counter-clockwise from the northern axis, the magnetic field vector is given by
\begin{equation}
\vec{B}=
\begin{pmatrix}
B_x\\
B_y\\
B_z
\end{pmatrix}
=
\begin{pmatrix}
B_{\rm sp} \cos(\theta+\phi)\\
B_{\rm ver}\\
B_{\rm sp} \sin(\theta+\phi)
\end{pmatrix},
\end{equation}
where $\phi$ is the pitch angle of the spiral field component and $B_{\rm sp}$ and $B_{\rm ver}$ are constants to scale the spiral and vertical field components. We obtain the observed magnetic field vectors $\vec{B}^{\rm obs}$ by rotation; the inclination $i$ corresponds to a rotation around the z-axis and the position angle, $pa$, corresponds to a rotation around the y-axis,
\begin{equation}
\vec{r}^{\rm obs}+\vec{B}^{\rm obs}=R_y(pa)\cdot R_z(i)\cdot (\vec{r}+\vec{B}),
\end{equation}
where $R_y$ and $R_z$ are the 3-D rotation matrices around the indicated axis by the inserted angle and $\vec{r}^{\rm obs}$ is the position seen by the observer. We note that $\vec{B}^{\rm obs}$ does not depend on the radius while for an inclined galaxy $B_x^{\rm obs}$, $B_y^{\rm obs}$ and $B_z^{\rm obs}$ vary with $\theta$ in this simplified model.

Our model does not include a radial dependence of $B_{\rm ver}$ or $B_{\rm sp}$, although it can easily be introduced. To compare the model with our data, we use sector averages in rings at fixed radii. Over a large part of the central disk, the radial changes should be small, and only in the innermost and outermost parts of the disk $B_{\rm ver}$ changes rapidly with $\vec{r}$. The effects of modifying the different parameters are discussed in Section~\ref{sec:robustness}. 

Figure~\ref{model_faceon} sketches the face-on view, where only the log-spiral field in the disk-plane is seen. 
Figure~\ref{model_edgeon} shows the vertical component seen edge-on. 
To mimic a divergence-free field, we introduce a radial dependence of $B_y$, so that the vertical field points towards the disk ($B_y<0$) in the inner disk, while it points away from the disk ($B_y>0$) in the outer disk. This sharp field reversal is a simplification of a quadrupolar/dipolar field. 
For comparison with our sector data (Section~\ref{sec:comparison}), we assume that we observe only the outer part of the central disk, where the magnetic field points away from the disk in Figure~\ref{model_edgeon}. This corresponds to the outer parts in Figs.~\ref{fig:dipol} or \ref{fig:quadrupol}.

Fig.~\ref{model_incl} shows both field components in the sky plane seen under the position angle of $70\grad$ and the inclination of $43\grad$ (see Sect.~\ref{sec:sector}). We observe the disk from below. At 6.2~cm, the far side is depolarized and thus only the vertical component of the near side is seen.
Due to this, the symmetry across the mid-plane is unknown, and we cannot distinguish between a dipolar and a quadrupolar field configuration. 
The observed magnetic field is the vector addition of the spiral component in the disk and the vertical component (we note that both are the components of the same quadrupolar/dipolar-type dynamo field), which is shown in Figure~\ref{model_incl_add}. The length of the vectors corresponds to the projected magnetic field strength in the sky plane and thus to the observed polarized intensity. The similarity to the observed polarized intensity at 6.2~cm (shown in contours) is obvious and shows that our model is valid in a qualitative way.

\subsection{Comparison with our sector data}\label{sec:comparison}

The observed polarized intensity is proportional to the projection of $\vec{B}^{\rm obs}$ to the xz-plane; the observed $RM$ is proportional to the y-component of $\vec{B}^{\rm obs}$, and the observed apparent pitch angle can be calculated from the projection of $\vec{B}^{\rm obs}$ to the xz-plane:
\begin{eqnarray}
PI = c_{pi} \left(\sqrt{(B_x^{\rm obs})^2+(B_z^{\rm obs})^2}\right)^{1-\alpha} \label{eq:PI}\\
RM = c_{rm} B_y^{\rm obs} + RM_{\rm fg} \label{eq:PA}\\
\phi_{\rm obs} = -\theta+\arctan\left(\frac{B_z^{\rm obs}}{B_x^{\rm obs}}\right)-pa, \label{eq:phi}
\end{eqnarray}
where $\alpha=-0.7$ is the spectral index (recall that the synchrotron intensity $I_\nu\propto B_\perp^{\,\,\,1-\alpha}\,\nu^\alpha$). The variable $RM_{\rm fg}$ is the rotation measure contribution from the foreground, which shifts the observed $RM$ pattern by a constant offset (see Sect.~\ref{sect:reversal}).

The variables $c_{pi}$ and $c_{rm}$ are proportionality constants used to scale the amplitude of the model output. In a physical context they hold information about the number density of relativistic particles per energy interval and the thermal electron density and path length along the line of sight. In this qualitative analysis we are only interested in the geometry of the magnetic field, which is fully determined by the proportionality between the observed quantities and the projected magnetic field components, not by their absolute quantities.

In Fig.~\ref{fig:model-on-data} we used $B_{\rm sp}=1.0$, $B_{\rm ver}=0.7$, $\phi=30.0\grad$, $c_{pi}=20.0$, $c_{rm}=150.0$, $i=43\grad$, and $pa=70\grad$ and plot the model output on top of the corresponding sector data (Section~\ref{sec:sector}) for $PI$, $RM$, and $\phi_{\rm obs}$. We note that we use the same model for all four rings (so a deviation for the innermost rings is expected) and we did not fit any parameters. We use four rings with a radial width of 0.1~kpc starting at 0.1~kpc radius from the centre. Our geometrical model describes the behaviour of all three observables well. 

\section{Robustness of the results}\label{sec:robustness}

\afterpage{%
    \clearpage
    \begin{figure*}[!hp]
    \begin{center}

    \includegraphics[trim=0px 0px 200px -30px,scale=0.4]{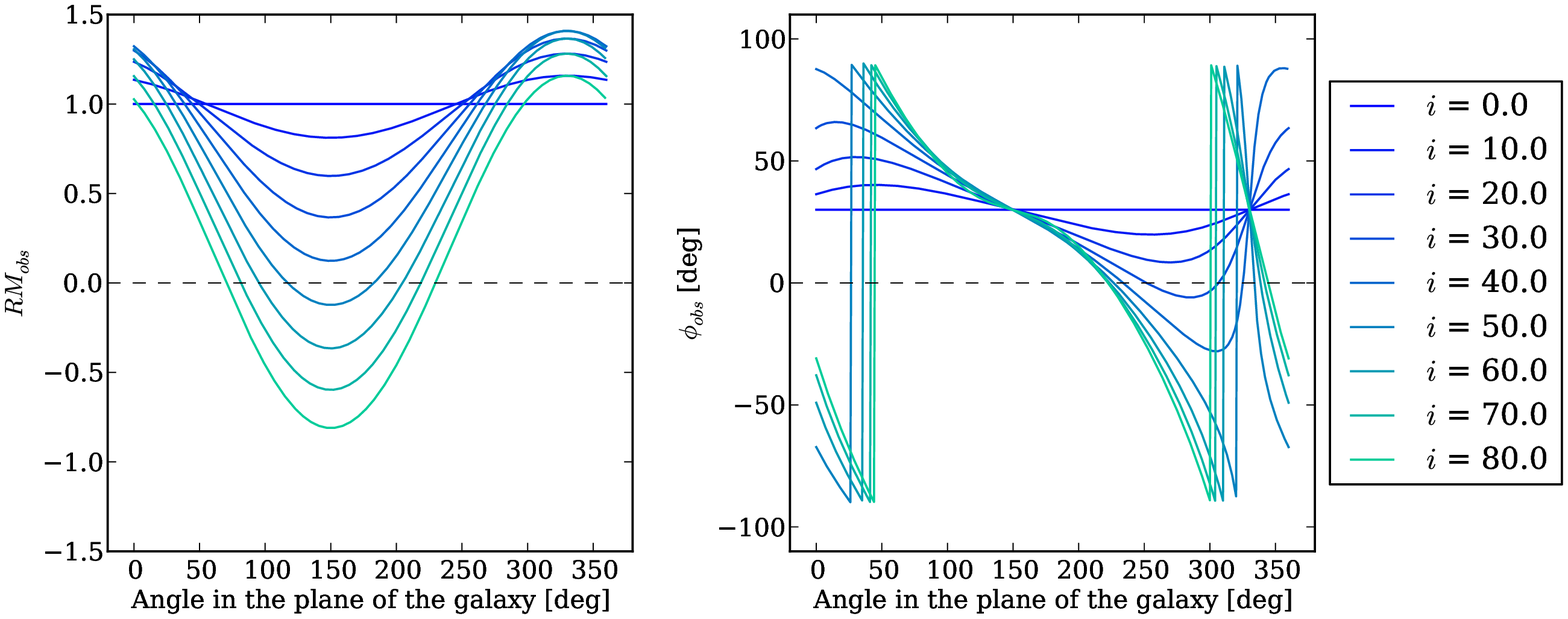}
    \caption{Model output (from left to right: polarized intensity, rotation measure, and observable pitch angle
    as a function of azimuthal angle) for different inclinations $i$.}
    \label{fig:var_inc}

    \includegraphics[trim=0px 0px 200px -30px,scale=0.4]{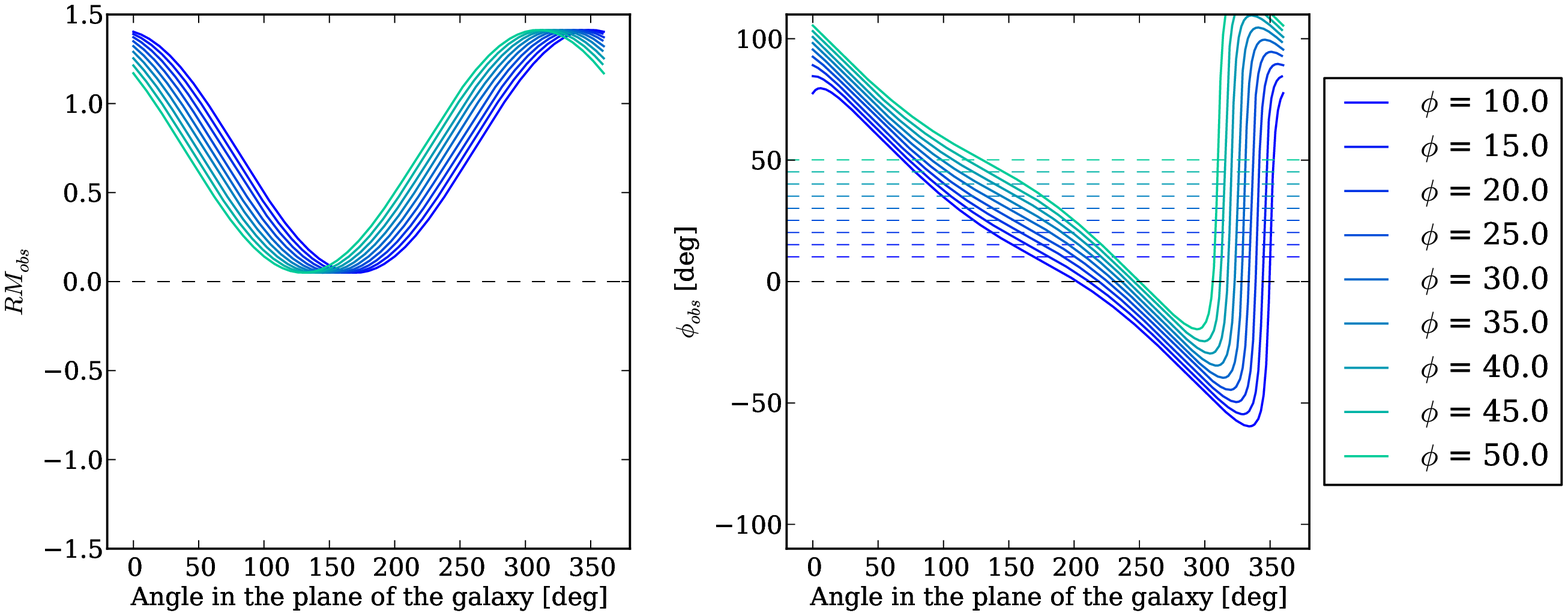}\vfill
    \caption{Model output (from left to right: polarized intensity, rotation measure, and observable pitch angle
    as a function of azimuthal angle) for different pitch angles $\phi$
    of the spiral field component. The dashed horizontal lines indicate the average values of $\phi_{\rm obs}$.}
    \label{fig:var_phi}

    \includegraphics[trim=0px 0px 200px -30px,scale=0.4]{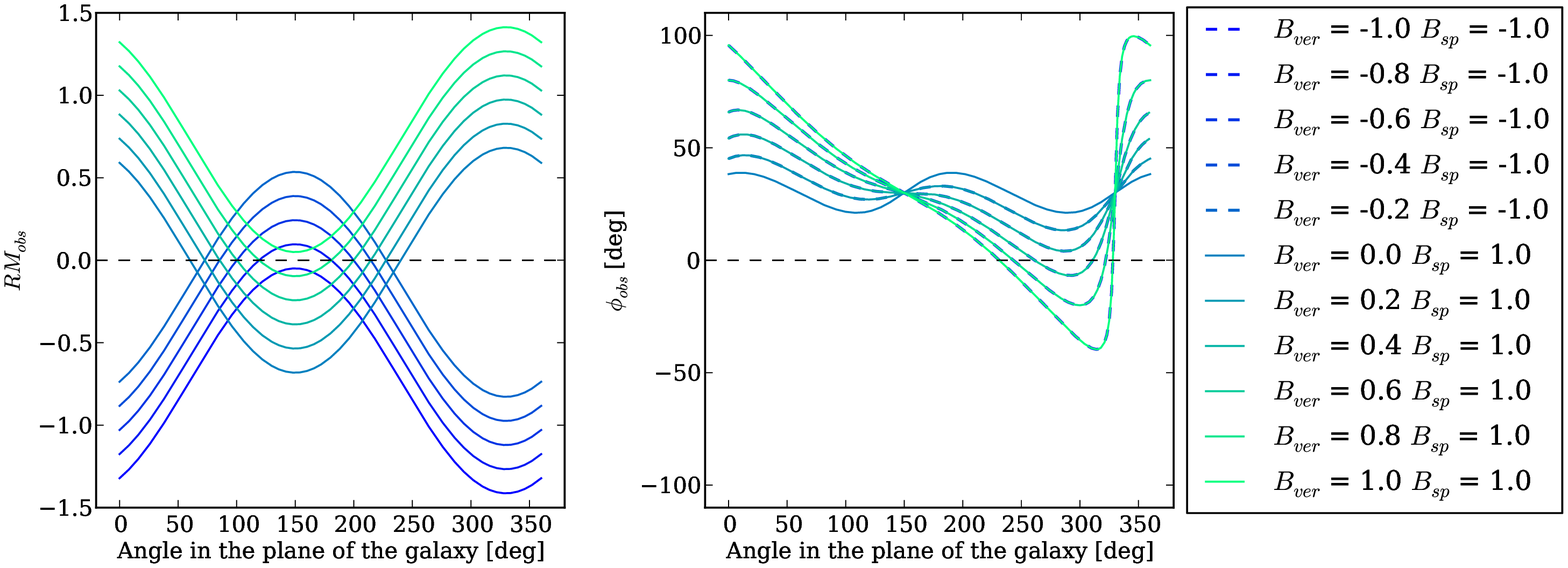}
    \caption{Model output (from left to right: polarized intensity, rotation measure, and observable pitch angle
    as a function of azimuthal angle) for different strengths of the vertical field component $B_{\rm ver}$.}
    \label{fig:var_By}

    \includegraphics[trim=0px 0px 200px -30px,scale=0.4]{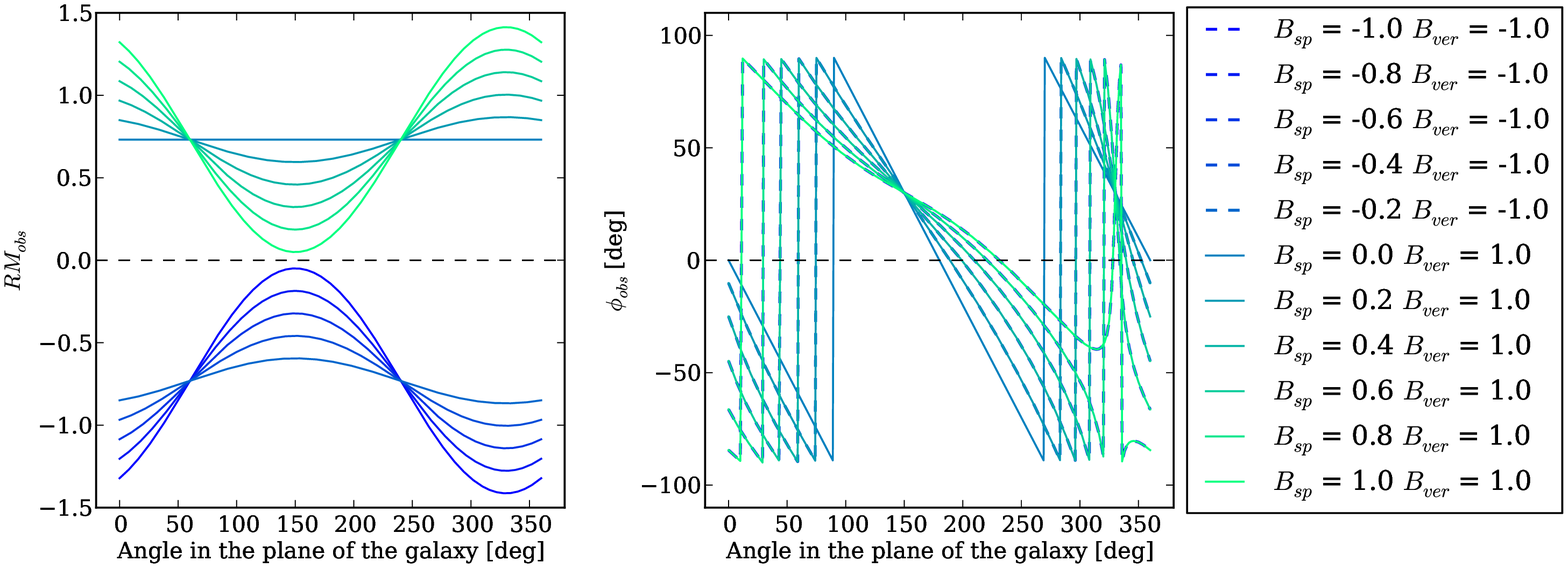}
    \caption{Model output (from left to right: polarized intensity, rotation measure, and observable pitch angle
    as a function of azimuthal angle) for different strengths of the spiral field component $B_{\rm sp}$.}
    \label{fig:var_Bs}
    \end{center}
    \end{figure*}

    \clearpage

    \begin{figure*}[!hp]
    \begin{center}
    \includegraphics[scale=0.55, bb=-75 -102 706 884]{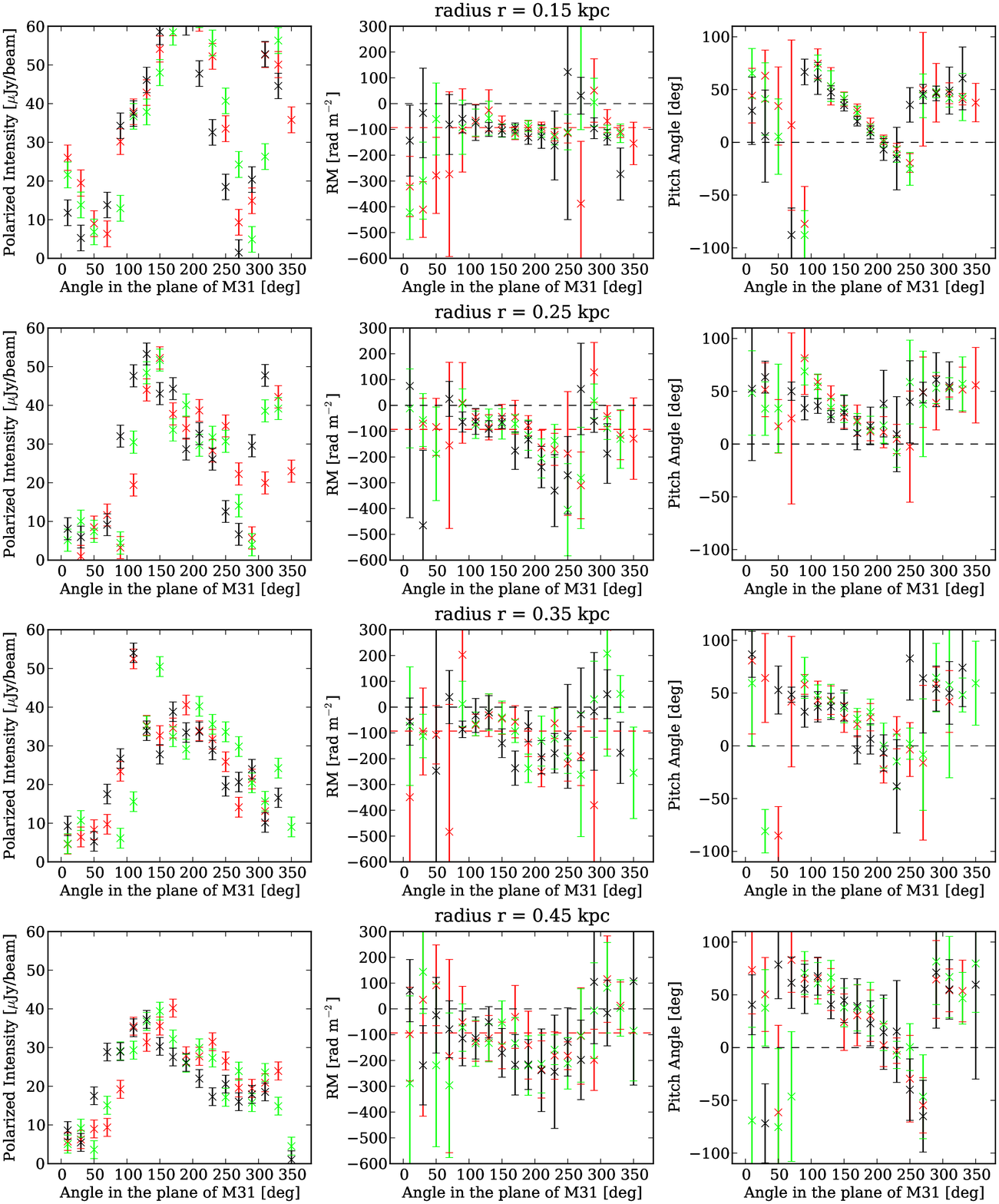}
    \caption{Result for different position angles of ellipses defining our sectors. Red: $pa=38\grad$ \citep[e.g.][]{boulesteix,ciardullo}, green: $pa=48\grad$ \citep{saglia}, and black: $pa=70\grad$ \citep{melchior}.}
    \label{fig:posangs}
    \end{center}
    \end{figure*}

    \clearpage
}

In Figs.~\ref{fig:var_inc} to \ref{fig:var_Bs}, we show the resulting model outputs for different parameters (see Equations~\ref{eq:PI} to \ref{eq:phi}), for comparison with Fig.~\ref{fig:model-on-data}. Only one parameter is varied at a time, the remaining parameters are fixed to $i=43\grad$, $\phi=33\grad$, $B_{\rm sp}=B_{\rm ver}=1.0$. The scaling parameters $c_{pi}$, $c_{rm}$ and $RM_{\rm fg}$ are not used (i.e. $c_{pi}=c_{rm}=1$, $RM_{\rm fg}=0$).

Figure~\ref{fig:var_inc} shows the behaviour for different inclinations. The variation in polarized intensity increases with inclination and starts to show two peaks when exceeding $40\grad$. The variations in $RM$ and $\phi_{\rm obs}$ also increase with inclination, so that a minimum inclination of $20\grad$ can be safely assumed from our data.
At least for the outer three rings, the data does not indicate a clear double-peak structure for $PI$ and the observed pitch angle does not significantly wrap around $\pm90\grad$. The behaviour of the observed $RM$ also excludes inclinations larger than $50\grad$, since the $RM$ does not change too rapidly with the azimuthal angle. The assumed inclination of $i=43\grad$ taken from the literature is consistent with our data but cannot be further constrained.

Varying the pitch angle of the spiral component $\phi$ does not change the observable patterns of $PI$, $RM$
and $\phi_{\rm obs}$ (Fig.~\ref{fig:var_phi}) but only causes a shift in azimuthal angle and $\phi_{\rm obs}$.
The mean of the observable pitch angle over the entire azimuthal range represents the pitch angle of the spiral field. 

The spiral pitch angle is thus well determined by our observations.

Figures~\ref{fig:var_By} and \ref{fig:var_Bs} illustrate that the direction of the field vectors is very well constrained by the three observables. The spiral field component $B_{\rm sp}$ must be positive (i.e. the field must point outwards) to explain the observed $RM$ pattern. The relative strength of $B_{\rm sp}$ and $B_{\rm ver}$ is best constrained by $\phi_{\rm obs}$. As expected for a quadrupolar or dipolar field configuration, our data is most consistent with $B_{\rm sp} \geq B_{\rm ver}$ and $B_{\rm ver}>0$.

\begin{figure}[htb]
\centering
\subfigure[Halo field (blue) of an antisymmetric dipolar magnetic field. The black arrows show the direction of the radial field component above and below the mid-plane.]{
    \includegraphics[scale=0.3]{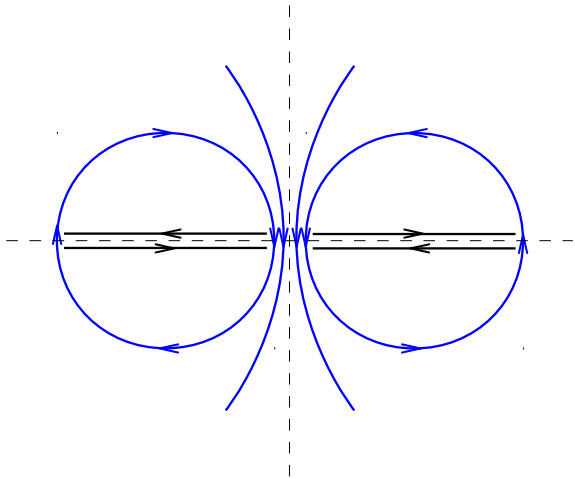}
    \label{fig:dipol}
}\vfill

\subfigure[Halo field (blue) of a symmetric quadrupolar magnetic field. The black arrows show the direction of the radial field component.]{
    \includegraphics[scale=0.3]{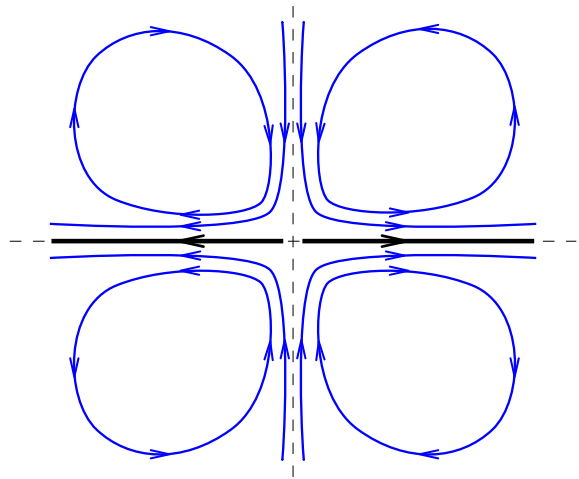}
    \label{fig:quadrupol}
}
\caption{Cartoon to visualize that the general directions of $B_{\rm sp}$ and $B_{\rm ver}$ are coupled. The poloidal component (blue) of a dipolar (\ref{fig:dipol}) and a quadrupolar (\ref{fig:quadrupol}) field. In addition, we show the direction of the radial field component (black). We note that for the dipolar field the radial component above and below the mid-plane are oppositely directed. The directions of the vertical and spiral components of our model are thus coupled.}
\label{fig:dipol-quadrupol}
\end{figure}

We note that the general directions of $B_{\rm sp}$ and $B_{\rm ver}$ cannot be chosen freely for a fixed radius, since the radial direction is coupled to the direction of the halo field. As illustrated in Fig.~\ref{fig:dipol-quadrupol}, if $B_{\rm ver}$ is directed outwards in the outer disk, the radial component has to point outwards as well and vice versa. This is obvious for a quadrupolar field. In a dipolar field, the radial component changes direction across the mid-plane. Our simple model can thus not distinguish between a dipolar or quadrupolar field configuration.

A combination of $B_{\rm sp}$ and $B_{\rm ver}$ with opposite sign is only possible in the innermost part, where the field lines return to the disk. If only $B_{\rm ver}$ changes its sign, the patterns of $PI$ and $\phi_{\rm obs}$ are shifted by $180\grad$ in azimuthal angle (which at least for $PI$ can be seen in Fig.~\ref{fig:model}). The deviations from the model in the innermost ring with radius 0.15 kpc (Fig.~\ref{fig:model-on-data}), especially for $PI$, give an indication for the closing of the field lines in the centre, but the number of independent resolution elements per sector does not allow a verification using the innermost part.

In Fig.~\ref{fig:posangs}, we plot the observed data using different position angles found in the literature to define our sectors, as this might affect the observed patterns. The general behaviour of $PI$, $RM$ and $\phi_{\rm obs}$ does not change for different position angles. Within the errors, it is not possible to further constrain the inclination nor the position angle of the central region with the present data.

\section{Conclusions}

(1) Our radio polarization maps of the central region of M31 at 3.6~cm and 6.2~cm, as obtained from combined data from the VLA and Effelsberg telescopes, reveal a spiral field with a pitch angle of $\sim33\grad$, which is much larger than the spiral pitch angle of $8\grad-19\grad$ in the outer disk. This difference can be explained in terms of the $\alpha-\Omega$ dynamo theory by the different scale heights ($\approx0.37$~kpc and $\approx1$~kpc, respectively) of the ionized gas.

(2) The central region is partly Faraday-thick at 6.2~cm, so that only emission from the near side is observed. The polarization degree at 6.2~cm is small around the north-eastern major axis of the projected disk and large on the south-western side.

(3) A three-dimensional field configuration can explain the observed patterns in polarized intensity, rotation measure and polarization angle in the inner kiloparsec of M31. With the present data and our simple model we can, however, not distinguish between a pattern resulting from the central part of an ASS--quadrupolar or ASS--dipolar configuration (which would both appear helical in the innermost part).

(4) The magnetic field vectors of the central magnetic field spiral in the radial range 0.2--0.5 kpc point outwards. This is a safe observational result from the rotation measures between 3.6~cm and 6.2~cm. As the magnetic field in the 10~kpc ring is known to be directed inwards \citep{fletcher04}, the inner and outer fields point into {\em opposite directions}. This result is not affected by the uncertainty in the orientation of the central disk. As the inner region and the outer disk are physically decoupled, we propose that two independent dynamo-active regions have generated these field patterns, as was already suggested by \cite{ruzmaikin81}, where the inner dynamo is more efficient and faster than the outer one.

\begin{acknowledgements}
Based on observations with the 100-m telescope of the MPIfR (Max-Planck-Institut für Radioastronomie) at Effelsberg.
All model calculations were made in \texttt{iPython} \citep{PER-GRA:2007} using \texttt{Matplotlib} \citep{Hunter:2007}.
The National Radio Astronomy Observatory is a facility of the National Science Foundation operated under cooperative agreement by Associated Universities, Inc. --
We thank Katia Ferri\`ere, Dmitry Sokoloff, Elly Berkhuijsen and our referee Andrew Fletcher
for very helpful discussions and comments. RB acknowledges support from DFG FOR1254. --
The presented results and images have in part been published in \citet{phd}.

\end{acknowledgements}

\bibliographystyle{aa} 
\bibliography{mybib} 

\end{document}